\documentclass[aps,prl,twocolumn,unsortedaddress]{revtex4-1}
\usepackage{amssymb}
\usepackage{graphicx}
\usepackage{floatflt}

\begin{document}

\title{Ultrafast dynamics of charge density waves in 4$H_{b}$-TaSe$_{2}$ probed by femtosecond electron diffraction }
\author{N. Erasmus$^{1,\ast}$, M. Eichberger$^{2,\ast}$, K. Haupt$^{1}$, I.
Boshoff$^{1}$, G. Kassier$^{1}$, R. Birmurske$^{2}$, H. Berger$^{3}$, J.
Demsar$^{2}$, and H. Schwoerer$^{1}$}
\affiliation{$^{1}$Laser Research Institute, Stellenbosch University, Stellenbosch 7600,
South Africa}
\affiliation{$^{2}$Department of Physics and Center of Applied Photonics, University of
Konstanz, 78464 Konstanz, Germany}
\affiliation{$^{3}$Physics Department, EPFL CH-1015 Lausanne, Switzerland}
\date{\today}

\begin{abstract}
The dynamics of the photoinduced commensurate to incommensurate charge density wave (CDW) phase transition in 4$H_{b}$-TaSe$_{2}$ are investigated by
femtosecond electron diffraction. In the perturbative regime the CDW reforms on a 150 ps timescale, which is two orders of magnitude slower than in other
transition-metal dichalcogenides. We attribute this to a weak coupling between the CDW carrying $T$ layers and thus demonstrate the importance of
three-dimensionality for the existence of CDWs. With increasing optical excitation the phase transition is achieved showing a second order character in contrast to the first order behavior in thermal equilibrium.
\end{abstract}


\maketitle

Reduced dimensionality seems to be a decisive property governing phenomena
like high temperature superconductivity and charge density wave (CDW)
formation. The latter is typical for quasi one- or two-dimensional metals
where, in the ground state, the crystal displays a static periodic modulation of the
conduction electron density accompanied by a periodic lattice displacement
(PLD), both characterized by the wave vector $\mathbf{q}%
_{\mathrm{CDW}}$ \cite{Gruner1985}. The standard theory that describes the
appearance of this macroscopic quantum state was formulated by Peierls
\cite{Peierls1956}. By considering a one dimensional metal he has shown that
the divergent static electronic susceptibility\ at a wave vector $\mathbf{q}%
=2\mathbf{k}_{\text{F}}$ gives rise to an instability of the electronic system against perturbations at
this wave vector. This so called Fermi surface nesting lowers the frequencies of $\mathbf{q}%
=2\mathbf{k}_{\text{F}}$ phonons, which will eventually evolve into a static lattice displacement. From that band gaps at $\pm$ $\mathbf{k}_{\text{F}}$ result, which reduce the total electronic energy. If the elastic energy cost
to modulate atomic positions is lower than the electronic energy gain, the CDW
state is the preferred ground state. Recently, this classical picture has been
challenged \cite{Johannes2008,Gorkov2012,Rossnagel2011}, since (a) the nesting
condition derived from the topology of the Fermi surface ($2\mathbf{k}%
_{\text{F}}$) and the observed CDW modulation vectors ($\mathbf{q}%
_{\mathrm{CDW}}$) are not generally equal, and (b) the diverging
susceptibility at $\mathbf{q}=2\mathbf{k}_{\text{F}}$ is exceedingly fragile
with respect to temperature, scattering or imperfect nesting
\cite{Johannes2008}. Contrasting the standard Fermi surface nesting scenario the
transition from the metallic into a CDW state was argued to occur due to strong
$\mathbf{q}$-dependent electron-phonon coupling \cite{Johannes2008},
particularly in transition-metal dichalcogenides \cite{Gorkov2012}.

Only the concerted interplay of electronic and lattice degrees of freedom make the CDW
formation possible. The two modulations can be individually examined by scanning
tunneling microscopy \cite{Burk1992}, angular resolved photoemission
spectroscopy \cite{Rossnagel2011} and electron, x-ray or neutron diffraction
techniques \cite{Luedecke1999}. In thermal equilibrium, the gap in the
electronic spectrum and the atomic displacement amplitude ($A$) present
different projections of the same order parameter \cite{Gruner1985}. Adding
femtosecond temporal resolution to the experiment enables investigation of
their dynamical behavior. Since the electronic system can be perturbed on
timescales much faster than the characteristic lattice vibration periods these
femtosecond real-time techniques enable studies of the interplay between these
two components of the order parameter. Studies of coherently driven collective
modes \cite{Demsar2002,Perfetti2006,Schaefer2010}, photoinduced phase
transitions \cite{Schmitt2008,Tomeljak2009,Rohwer2011,Petersen2011} as well as
manipulation of the order parameter \cite{Mihailovic2002,Yusupov2010} have
been performed with femtosecond resolution in both quasi one- and
two-dimensional CDW compounds. By tracking the photoinduced changes in optical
properties or in photoelectron emission spectra, information about the transient changes in
the electronic subsystem of the CDW can be obtained, while the lattice
dynamics can be deduced only indirectly \cite{Schaefer2010}. With the
recent development of time-resolved structural probes, such as ultrafast x-ray
\cite{Sokolowski2003,Fritz2007,Beaud2009} and electron diffraction
\cite{Siwick2003,Baum2007c,Sciaini2011} techniques, structural dynamics in
solids are becoming experimentally accessible
\cite{Eichberger2010,Vorobeva2011}, enabling direct insight into the coupling
strength between electrons and the lattice and their dynamical interplay.

In this Letter we report on systematic excitation density studies of
structural dynamics in the CDW system 4$H_{b}$-TaSe$_{2}$ in its room
temperature phase. Here commensurate (C) CDWs exist in the octahedrally coordinated layers. Upon strong enough excitation the phase transition to an incommensurate (IC) high temperature CDW phase can be achieved on the sub picosecond timescale.
By means of femtosecond electron diffraction we were able to directly probe
the structural order parameter dynamics following photoexcitation with
near-infrared femtosecond optical pulses. Resolving the CDW suppression
dynamics on the sub picosecond timescale we demonstrate the importance of
specific strongly coupled phonons for the disappearance of the CDW. Moreover, the CDW
reformation dynamics is found to be dramatically slower than in recently
studied 1$T$-TaS$_{2}$ \cite{Eichberger2010} and 1$T$-TiSe$_{2}$
\cite{Vorobeva2011} implying the importance of three-dimensionality for the
existence of CDW order. Finally, by studying the non-equilibrium CDW state
generated by optical perturbation as a function of excitation density we
demonstrate the change of order of the C-IC phase transition: while in thermal equilibrium it is of strongly first order the ultrafast optically induced
phase transition is of second order type.

\begin{figure}[tbp]
\vspace{2ex} \includegraphics[width=0.9\columnwidth]{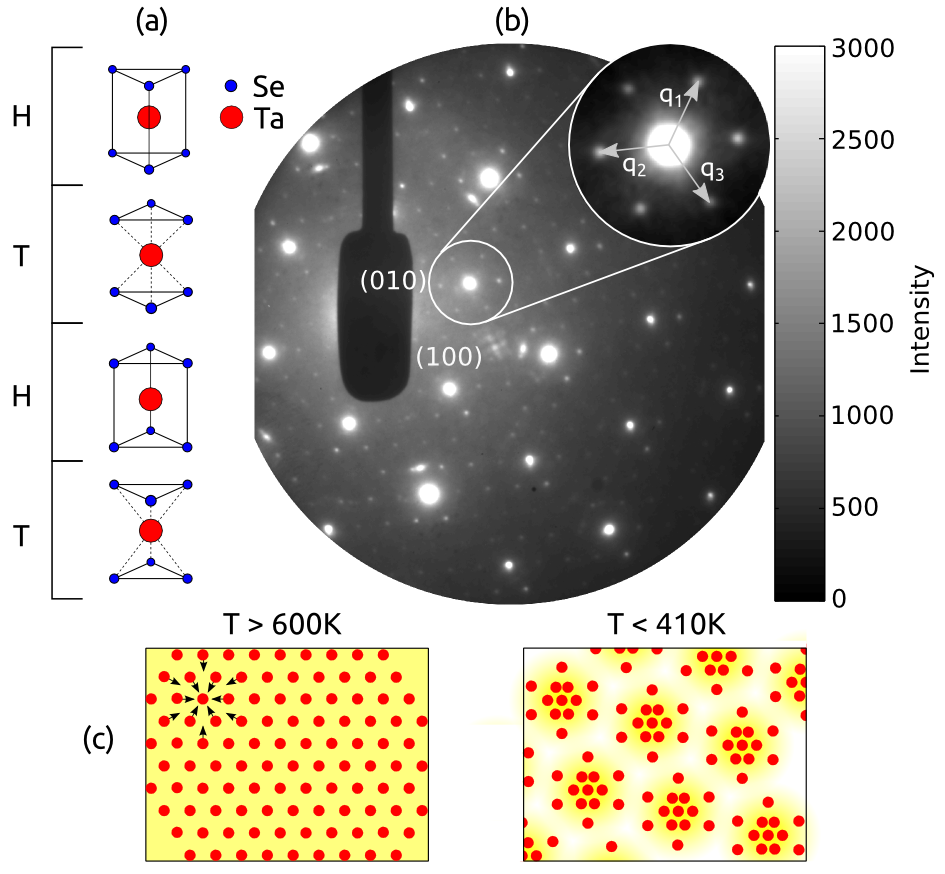}
\caption{{Crystal structure and charge density waves in 4$H_{b}$-TaSe$_{2}$: Panel (a) depicts the alternating stacking of $T$ and $H$ layers of the unit cell. The equilibrium diffraction pattern of the studied sample is shown in panel (b). Here two primitive lattice points are indexed. The insert shows the six CDW satellites around each Bragg peak. Panel (c): above 600 K (metallic phase) the Ta ions are hexagonally arranged. Below $T_{\text{C}}=410\,$K the Ta ions of the $T$ layers are commensurately  ($\protect\sqrt{13}\times \protect\sqrt{13})$ modulated: twelve ions cluster around one central ion to form a commensurate PLD. The conduction electron density (shaded) is modulated accordingly.}}
\label{fig:diffpatt}
\end{figure}

4$H_{b}$-TaSe$_{2}$ is a quasi two-dimensional crystal consisting of
covalently bound three atom thick planar layers of type Se-Ta-Se which are
weakly bound to each other along the crystallographic $c$-axis via van der
Waals interaction, see Fig.\thinspace\ref{fig:diffpatt}(a). In the high
temperature metallic phase 4$H_{b}$-TaSe$_{2}$ belongs to the space group
P6$_{3}$/mmc with the hexagonal lattice parameters of $a=b=3.455\,{\mathring
{A}}$ and $c=25.15\,{\mathring{A}}$ \cite{Brown1965}. Here, octahedrally
coordinated $T$ layers alternate with trigonal prismatic $H$ layers, with the
individual Ta atoms being aligned on top of each other along the $c$-axis.
This structure serves as a host lattice for CDW induced modulations at lower
temperatures: below 600 K an IC-CDW in the $T$ layers is
formed with $\mathbf{q}_{\mathrm{CDW}}=0.265$ $\mathbf{a}^{\ast}$. By further
lowering the temperature through a first order phase transition at
$T_{\text{c}}=410\,$K a C-CDW with $\mathbf{q}_{\mathrm{CDW}}=0.277$
$\mathbf{a}^{\ast}$ ($\mathbf{q}_{\text{CDW}}$ is tilted by 13.9$^{o}$ from $\mathbf{a}^{\ast
}$) is formed \cite{Wilson1975,Salvo1976}. The periodic lattice
displacement of the charge density wave appears as ($\sqrt{13}\times\sqrt
{13})R13.9$ modulation, with no modulation vector along the $c$-axis
\cite{Luedecke1999}. The modulation, schematically shown in Figure
\ref{fig:diffpatt} (c), is reflected in the electron diffraction pattern. Here
each of the Bragg peaks of the host lattice is surrounded by six weak
satellite reflections whose intensity is only a few percent of the intensity
of the Bragg peaks of the host lattice, see Fig.\thinspace\ref{fig:diffpatt}%
(b). In both phases, C and IC, no CDW order is
present in the $H$ layer; a slight modulation therein can be assigned to
elastic coupling between the two types of layers \cite{Luedecke1999}. Only
upon further cooling below 75$\,$K, the $H$ layers develop an IC-CDW
which coexists with the still commensurately modulated $T$ layers
\cite{Wilson1975,Luedecke1999}. Thus, the CDW in 4$H_{b}$ -TaSe$_{2}$ at room
temperature presents a good approximation to a true two-dimensional CDW system.

Single crystals of 4$H_{b}$-TaSe$_{2}$ were grown by chemical vapour
transport. From the bulk crystal 90$\,$nm thin sheets $>150\,\mu$m in diameter
were cut along the TaSe$_{2}$ planes using an ultramicrotome and mounted on
TEM meshes \cite{Eichberger2012}. Time-resolved electron diffraction
experiments were performed in transmission geometry in a collinear excitation
scheme applying $200\,$fs near-infrared pump pulses ($\lambda=775\,$nm) for
photoexcitation, and delayed ultrashort electron pulses as probe. We used a
30$\,$kV femtosecond electron gun with a gold photocathode, driven by
150$\,$fs laser pulses ($\lambda=258\,$nm) at a repetition rate of 1$\,$kHz to
produce electron bunches containing $\sim1000$ electrons. Their duration at
the sample position was independently measured by a compact ultrafast streak
camera to be 300$\,$fs \cite{Kassier2010}. The lateral FWHM of the electron
beam on the sample was 200$\,\mu$m, while its transverse coherence length was
measured to be 5$\,$nm. Variable time delay $t_{\text{d}}$ between optical pump pulse
and the electron probe pulse was realized with a computer controlled
translation stage in the pump laser path. Accurate spatial overlap of electron
and laser pulses on the sample was actively controlled by monitoring the
position of the electron and laser spot and correcting the overlap with piezo
driven mirrors if necessary. Diffraction patterns were detected with a pair of
chevron stacked micro channel plates coupled to a phosphor screen, and imaged
onto a 16 bit CCD camera. Diffraction images at different time delays were
taken with 60 seconds exposure (60,000 shots). Each image with the pump laser
pulse incident on the sample was followed by an identical unpumped exposure
for reference. The maximum achievable temporal resolution was about 400$\,$fs,
if $\lesssim1000$ electrons per bunch were used. At large time-delays the data
were recorded with higher numbers of electrons per pulse to increase the
signal to noise ratio and decrease the acquisition time.

We have performed the experiments at room temperature and analyzed the temporal 
evolution of photoinduced changes in the
intensity of the Bragg peaks and their CDW satellites at the individual
diffraction orders. In all six diffraction orders both satellite and Bragg
peaks showed identical temporal evolution. Thus, for the sake of increasing
the signal to noise level, individual CDW and Bragg traces were averaged over
all available diffraction orders \cite{Eichberger2010}. The experiments were
performed as a function of excitation fluence $F$, which was varied between 1
and 3 mJ/cm$^{2}$. Since the optical beam diameter was 500 $\mu$m and the
optical penetration depth of 110 nm \cite{Beal1978} is larger than the film
thickness nearly homogenous excitation is achieved.

\begin{figure}[tbp]
\includegraphics[width=0.9\columnwidth]{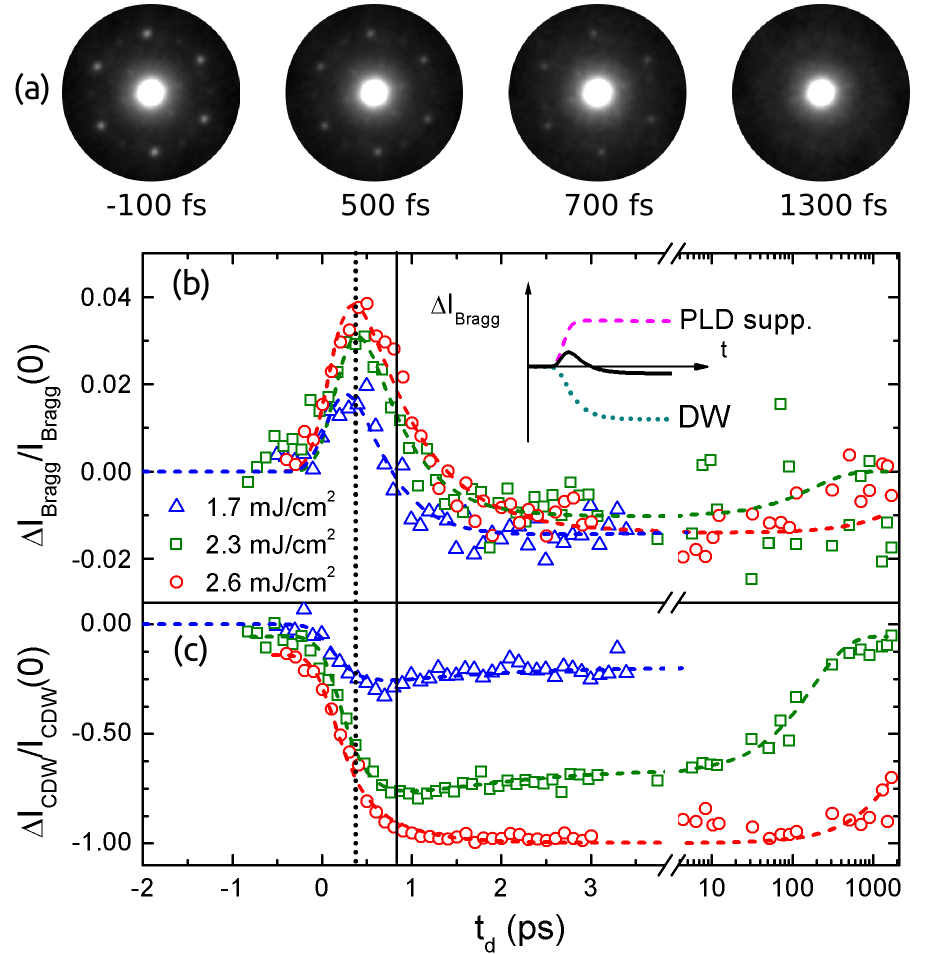}
\caption{{Temporal evolution of the diffraction intensities. Panel (a) shows the evolution of the insert in Fig.\ref{fig:diffpatt}(b) at several pump probe delays. Panels (b) and (c) represent the transient intensity changes in the Bragg and CDW reflections at different fluences $F$, respectively. The dashed lines are fits to the data, see text. The vertical solid and dashed lines indicate the times of the minimum of  $I_{\rm CDW} $ and the maximum of $I_{\rm Bragg}$ at $F=2.3$ mJ/cm${{}^{2}}$, respectively. The inset to panel (b) shows the two exponential components which in superposition (solid black) give the fit curve for the Bragg transient.}}
\label{fig:transients}
\end{figure}

Four diffraction images at different time-delays at \mbox{$F=2.6$ mJ/cm$^{2}$} are
shown in Fig. 2(a). At this excitation density the intensity of the CDW peaks
of the C-phase is completely suppressed on a sub-ps timescale, corresponding
to the photoinduced C-IC phase transition. Due to the reduced PLD amplitude in
the IC-phase the weak CDW peaks of the IC-phase are not resolved. Figure
\ref{fig:transients} (b) and (c) show the evolution of the changes in Bragg peak
($I_{\text{\textrm{Bragg}}}$) and CDW satellite ($I_{\text{\textrm{CDW}}}$)
intensities for three different excitation fluences as a function
of time-delay $t_{\text{d}}$.\ The suppression of $I_{\text{\textrm{CDW}}}$
ranges from $\approx30\,\%$ at $F=1.7$ mJ/cm$^{2}$ to a complete suppression
of the C-CDW order at $F=2.6$ mJ/cm$^{2}$. As the CDW is partially
suppressed, $I_{\text{\textrm{Bragg}}}$ gains in intensity as the crystal is
driven towards a more symmetric state, see Fig.\thinspace2(b). This process is
followed by a rapid decrease of $I_{\text{\textrm{Bragg}}}$ accompanied by a
further reduction in $I_{\text{\textrm{CDW}}}$, which take place on the sub-ps
timescale. For the perturbative regime, at $F<2.5$ mJ/cm$^{2}$, the recovery
of the CDW state is observed on the 150 ps timescale, while for $F>2.5$
mJ/cm$^{2}$ the recovery proceeds on a much longer timescale.

In order to analyze the structural dynamics we propose the following simple
model. Photoexcitation of electrons and initial e-e scattering result in pronounced changes
also in the electron distribution of the Ta 5d band. This process gives rise to
strong modification of the interatomic potential thereby launching highly
cooperative (coherent) atomic motion towards the new equilibrium positions of
the high-T phase via the breathing amplitude mode of the 13 Ta atom clusters
in Fig. \ref{fig:diffpatt}\thinspace(c). Thus, the PLD amplitude is suppressed
as manifested by the suppression of $I_{\mathrm{CDW}}$ and the corresponding
increase in $I_{\text{\textrm{Bragg}}}$. This timescale, $\tau
_{\text{\textrm{coh}}}$, is found to be shorter than our experimental
time-resolution. Following the transient increase $I_{\text{\textrm{Bragg}}}$
starts to decrease on a timescale $\tau_{\text{\textrm{icoh}}}$, during which
$I_{\text{\textrm{CDW}}}$ gets further suppressed. This timescale is
characteristic of electron-phonon relaxation processes, whereby the excited
$\mathbf{q}\neq0$ phonons give rise to suppression of $I_{\text{\textrm{Bragg}%
}}$ similar to the Debye-Waller effect. Similarly, the phonons which are
strongly coupled to the Ta 5d electrons seem to further destabilize the CDW.
Thus, we modeled the relative changes of Bragg and CDW scattering intensities
as $C_1(1-\exp(-t_{\text{d}}/\tau_{\text{\textrm{coh}}}))+C_2(1-\exp(-t_{\text{d}}%
/\tau_{\text{\textrm{icoh}}}))$, with $C_1,C_2<0$ for the case of $\Delta
I_{\text{\textrm{CDW}}}\left(  t_{\text{d}}\right)  $ and $C_1>0$, $C_2<0$ for $\Delta
I_{\text{\textrm{Bragg}}}\left(  t_{\text{d}}\right)  $; the latter has been sketched
for clarity in inset to Figure \ref{fig:transients}\thinspace (b).

Three modes around$~70\,$cm$^{-1}$ have previously been reported to show the
characteristic softening of CDW amplitude modes upon warming towards
$T_{\text{c}}$ \cite{Nakashizu1984}. Since the ultimate timescale for the
(coherent) suppression of the C-CDW is given by 1/4 of a period of
the amplitude mode we chose $\tau_{\text{\textrm{coh}}}=150$ fs. Moreover, to
fit the entire traces, the function has been multiplied by an exponential
decay to account for the reformation dynamics, and convoluted with a Gaussian
pulse with FWHM of 400\thinspace fs to make up for the temporal resolution.
From the fit with this simple model (dashed lines in Fig.\thinspace2) we
obtain $\tau_{\text{\textrm{icoh}}}\approx500\,$fs. Indeed, these two
processes seem to be required to consistently describe both $\Delta
I_{\text{\textrm{CDW}}}\left(  t_{d}\right)  $ and $\Delta I_{\mathrm{Bragg}%
}\left(  t_{d}\right)  $. It is due to $\tau_{\text{\textrm{coh}}}%
<\tau_{\text{\textrm{icoh}}}$ that $I_{\text{\textrm{CDW}}}$ reaches its
minimum several 100 fs later than the maximum in $I_{\text{\textrm{Bragg}}}$,
see solid and dashed vertical lines in Figure \ref{fig:transients} (b) and (c).
Interestingly, this delay is becoming larger with increasing $F$, which can be
attributed to different $F$-dependence of the two processes.

\begin{figure}[tbp]
\includegraphics[width=0.9\columnwidth]{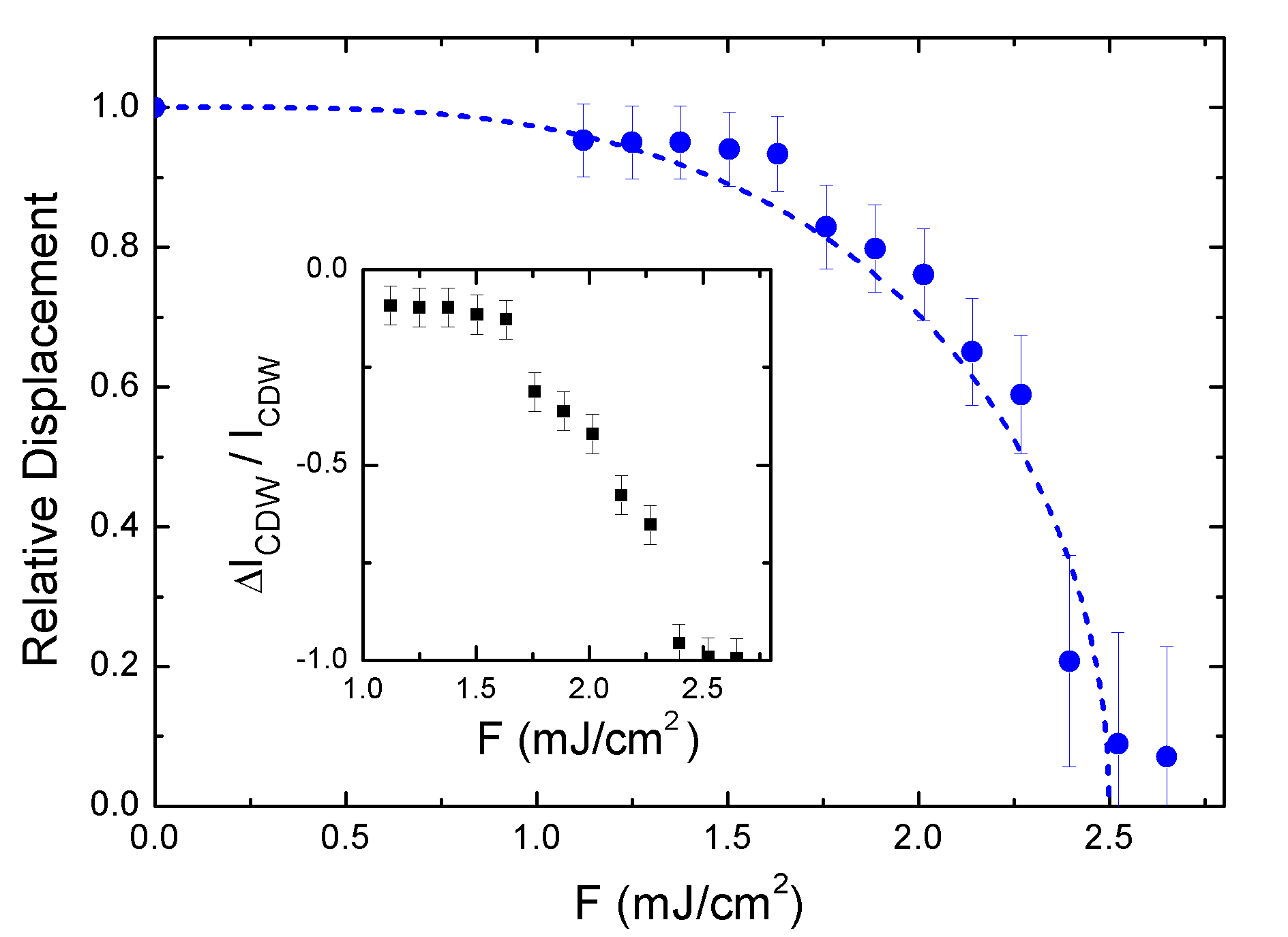}
\caption{Fluence dependence of transient PLD displacement at $t_{d}=1$ ps relative to its unperturbed value. The data are fit with the BCS function, where temperature was replaced by fluence as control parameter. The insert shows the corresponding changes in $I_{\rm CDW}$.}
\label{fig:fluence}
\end{figure}

The observation that the complete C-IC phase transition is achieved only after
a substantial part of the absorbed energy is transferred to lattice vibrations
via e-ph relaxation suggests that the optically induced phase transition
relies not only on launching the amplitude modes via electronically driven change of the interatomic potential, but also on exciting specific
strongly coupled lattice modes. Thus, the photoinduced phase transition seems
not to be purely electronically driven. Further support for the importance of
strong e-ph coupling comes from the fact that the optically delivered energy
required to fully drive the C-IC phase transition corresponds well to the
energy required to heat up the sample to $T_{\text{c}}=410\,$K. Using the
measured absorbed energy and the specific heat of $c_{\text{p}}=85.7\,$%
\thinspace J/(mole K) \cite{Bolgar1992} we indeed obtain a temperature
increase of $\sim110\,$K at the critical fluence $F_{\text{c}}=2.5\,$%
mJ/cm$^{2}$, see Fig.\thinspace\ref{fig:fluence}. This finding is in agreement
with previous studies on a quasi-1D CDW in K$_{0.3}$MoO$_{3}$
\cite{Tomeljak2009} and 1$T$-TaS$_{2}$ \cite{Eichberger2010}, while in
$1T$-TiSe$_{2}$ the absorbed optical energy density required to drive the
phase transition was reported to be lower than the corresponding thermal one
\cite{Vorobeva2011}.

In the perturbative regime at $F<2.5$ mJ/cm$^{2}$ the CDW reformation is found
to proceed with the timescale $\tau_{\mathrm{rec}}\approx150$ ps, see
Fig.\thinspace\ref{fig:transients}\thinspace(c). $\tau_{\mathrm{rec}}$ is about
two orders of magnitude larger than the observed recovery times in
$1T$-TaS$_{2}$ ($\approx4$\thinspace ps) \cite{Eichberger2010} and
$1T$-TiSe$_{2}$ ($\approx1$\thinspace ps) \cite{Vorobeva2011}. Thus, the CDW
in 4$H_{b}$-TaSe$_{2}$ seems far less cooperative than in the two above
mentioned dichalcogenides. The most obvious difference between the polytypes
is the interplane coupling between the $T$ layers, which is strongly reduced
in 4$H_{b}$-TaSe$_{2}$ due to the intermediate $H$ layers which are not
contributing to the CDW.
The reduced coupling in 4$H_{b}$-TaSe$_{2}$ is demonstrated also by the
reduction of the phase transition temperature from $T_{\text{c}}=473\,$K for
1$T$-TaSe$_{2}$ to $T_{\text{c}}=410\,$K \cite{Salvo1976}. In this respect the
dramatically prolonged CDW recovery in 4$H_{b}$-TaSe$_{2}$ as opposed to the
other two 1$T$ dichalcogenides hints to the importance of a certain degree of
three-dimensionality for the formation of CDW order in these systems. In other
words, these results suggest that a purely two dimensional CDW order, as
e.g.\thinspace in a single $T$-TaSe$_{2}$ layer, might not be possible.

In the regime above $F>2.5$ mJ/cm$^{2}$ where the CDW is completely
suppressed, the CDW recovery proceeds on a much longer ($>$ ns) timescale. For
this time delays the system can be described simply by an elevated
temperature. Cooling is governed by the heat diffusion, which is slow due to
the weak thermal coupling of the free standing film to the copper mesh.

To gain additional insights into the nature of the transient state, we
performed a detailed study of the structural order parameter (the displacement
amplitude) as a function of the excitation density. The displacement amplitude
relative to the unperturbed state is extracted directly from the relative
change in $I_{\text{\textrm{CDW}}}$ (inset to Fig. \ref{fig:fluence}), taking
into account that $I_{\text{\textrm{CDW}}}\propto(\mathbf{Q.A})^{2}$, where
$\mathbf{Q}$ is the CDW wave vector and $\mathbf{A}$ is the atomic
displacement.

In thermal equilibrium the C-IC phase transition has been identified to be of
a strong first order \cite{Salvo1976}. This is consistent with the
$F$-dependence of $I_{\text{\textrm{CDW}}}$ in a quasi-thermal equilibrium at
t$_{d}\approx1$ ns, where a step like change between $F=2.3$ and $2.6$
mJ/cm$^{2}$ is observed - see Fig. 2(c). The functional behavior of the order
parameter at t$_{d}=1$ ps upon increasing $F$ (Fig. \ref{fig:fluence}) clearly
shows a different character. Indeed, the order parameter ($\left\vert
\mathbf{A}\right\vert $\textbf{)} at t$_{d}\approx1$ ps can be very well
fitted by the BCS functional form, characteristic for second order phase
transitions. The dashed line in Fig. \ref{fig:fluence} is given by a well
known analytic approximation to the BCS gap equation \cite{Meservey1969}: here
instead of temperature $T$ fluence $F$ is used as a control parameter,
$\Delta/\Delta_{0}=\tanh(1.78(F_{c}/F-1)^{1/2})$, with the critical fluence
$F_{c}=2.5\,$mJ/cm$^{2}$. This clearly suggests a non-thermal character of the
photoinduced state. Moreover, this observation could present an important
clue for the theoretical understanding of interplay between electrons and lattice leading to the appearence
of CDWs in layered dichalcogenides.

In conclusion, we have studied the structural order parameter dynamics in the
CDW compound 4$H_{b}$-TaSe$_{2}$ using femtosecond electron diffraction. We
show that the CDW is suppressed on the sub-ps timescale, launched by the
electronically driven changes of the interatomic potential, and further
enhanced by strongly coupled lattice modes generated via e-ph scattering. In
the perturbative regime the CDW recovery time in 4$H_{b}$-TaSe$_{2}$
($\approx150$ ps) is found to be two orders of magnitude larger than
previously observed structural recovery times in similar transition-metal
dichalcogenides. This observation highlights the importance of
three-dimensionality for the existence of CDWs in layered compounds. Finally,
the fluence dependence of the CDW order parameter showed, that the
photoinduced ultrafast C-IC phase transition in 4$H_{b}$-TaSe$_{2}$ is of
second order in contrast to a first order thermally driven (slow) transition.

This work is based upon research supported by the South African Research Chair
Initiative of the Department of Science and Technology and the National
Research Foundation, the Alexander von Humboldt Foundation and Center for
Applied Photonics at University of Konstanz. M. E. gratefully acknowledges a
scholarship from Stiftung der Deutschen Wirtschaft (sdw). We acknowledge
valuable discussions with A.S. Alexandrov, V.V. Kabanov, I.I. Mazin, K. Rossnagel and S. van Smaalen.
\par
$^{\ast}$These authors contributed equally to this work.

\bibliographystyle{apsrev4-1}
\bibliography{ErasmusBib}

\end{document}